# New Approaches to Target Mass Corrections

F. M Steffens

*Universidade Presbiteriana Mackenzie*
*Rua da Consolação 896*
*01302-907*
*São Paulo – SP – Brasil*

**Abstract.** We analyze different prescriptions for the inclusion of target mass effects in the extraction of parton distributions from the measured structure functions. As a main result, the problem of defining parton distributions in the presence of mass is an open problem.

## INTRODUCTION

In the last few years, experiments have advanced to the region of large Bjorken $x_B$ [1]. This feature implies that we need to have a better control on the use of standard parton distributions in this region, which includes higher twist contributions and corrections coming from a finite value for the mass of the target. In the present talk, we will analyze different prescriptions for the inclusion of target mass effects and the meaning of a parton distribution in such case.

We start by defining the variables that will be used in our calculations. They are:

$$P^\mu = p^\mu + \tfrac{1}{2} m_N^2 n^\mu \qquad (1)$$

$$q^\mu = -\xi p^\mu + \frac{Q^2}{2\xi} n^\mu \qquad (2)$$

$$k^\mu = x p^\mu + \frac{k^2 + k_T^2}{2x} n^\mu + k_{T\mu} \qquad (3)$$

$$x_B = \frac{Q^2}{2P\cdot q}; \; \xi = \frac{Q^2}{2p\cdot q}; \; x = k\cdot n \qquad (4)$$

$$p^2 = n^2 = p\cdot k_T = n\cdot k_T = 0; \; p\cdot n = 1 \qquad (5)$$

In these equations, $P^\mu$ is the target four vector, $q^\mu$ the photon four vector, $k^\mu$ the parton four vector, $\xi$ the Nachtman variable, $x$ the parton momentum fraction, and $p^\mu, n^\mu$ auxiliary four vectors.

We will address four different prescriptions for the inclusion of target mass corrections. They are: Georgi and Politzer [2], D'Alesio, Leader and Murgia (DLM) [3], Steffens and Melnitchouk [4] and Accardi and Qui [5].

# TARGET MASS IN DIFFERENT PRESCRIPTIONS

We start with the handbag diagram, where the parton with momentum $k$ is hit by a photon with momentum $q$. If the partons are always on mass shell, then

$$\delta[(k+q)^2] = \frac{1}{2P\cdot q}\delta\left(-x_B + x + x\xi x_B \frac{m_N^2}{Q^2}\right), \qquad (6)$$

with $k_T^2 = 0$. In the Bjorken limit, Eq. (6) implies that $x = x_B$. In general, however, one has:

$$\frac{Q^2}{m_N^2} = \frac{x_B \xi^2}{x_B - \xi} \Rightarrow x = \xi. \qquad (7)$$

This result implies that for any finite $Q^2$, the Nachtman variable should be the correct variable to use for the parton momentum fraction.

## Georgi and Politzer Approach – The OPE

This is the approach based on the operator product expansion where in the product of currents terms proportional to $m_N^2/Q^2$ are kept. As a result, the $F_2$ structure function is written as:

$$F_2(x_B, Q^2) = \frac{\xi^2(1-a^2\xi^2)}{(1+a^2\xi^2)^3}F(\xi) + 6a^2\frac{\xi^3(1-a^2\xi^2)}{(1+a^2\xi^2)^4}H(\xi) + 12a^4\frac{\xi^4(1-a^2\xi^2)}{(1+a^2\xi^2)^5}G(\xi) \qquad (8)$$

where $a^2 = m_N^2/Q^2$, $A_n = \int_0^1 dx\, x^n f(x)$, $H(\xi) = \int_\xi^1 dy\, f(y)$ and $G(\xi) = \int_\xi^1 dy\, H(y)$. The parton distribution is $f(x)$.

The problem with this prescription is that it uses parton distributions in an unphysical region. This happens because in the presence of a target mass, the distributions are not defined up to 1 in the momentum fraction. Instead, the maximum momentum fraction is:

$$\xi_0 \equiv \xi(x_B = 1) = \frac{2}{1+\sqrt{1+4a^2}} \qquad (9)$$

which is smaller than 1 for any finite $Q^2$.

## D'Alesio, Leader and Murgia Approach

They follow closely the work of Ellis, Furmanski and Petronzio [6], where a parton model picture, including parton transverse momentum, is developed. In the case of DLM, they keep partons on mass shell, $k^2 = 0$, but retain they transverse momentum.

They call it the transverse basis. Their calculation reproduces the OPE results, including the problem with the upper limit in the integrals for the moments of the parton distributions implying, as before, contribution from an unphysical region.

## Steffens and Melnitchouk

Here, the definition of the moments of the parton distributions is modified in order to have integration over the physical region only:

$$A_n = \int_0^{y_0} dy\, y^n f(y), \qquad (10)$$

where $y_0$ is the maximum value physically supported by the distribution $f(y)$. Repeating the steps of Georgi and Politzer, they get $y_0 = \xi_0$, implying that now only physical contributions enter in the calculation of the target mass corrections to the structure functions. On the other hand, as seen from Eq. (9), the moments of the parton distribution depend on which target is being used. In other words, they are no longer universal and become $Q^2$ dependent! We are left with a true conundrum: If unphysical contributions are allowed, the partonic interpretation is retained. However, if one uses only the physical region, one loses the partonic interpretation.

## Accardi and Qiu

They use the collinear factorization in the impulse approximation and assume that in the graph of Figure 1 the relevant contribution comes from the lower part of the graph, where they impose that no baryon number flows through the upper part.

The main differences to the DLM approach are: (a) the parton momentum, in the calculation of the kinematics, is assumed to have no transverse momentum, and (b) the parton in the upper part of the handbag diagram can be off-mass shell. Thus,

$$0 \leq (k+q)^2 \leq (P+q)^2 - m_N^2 \Rightarrow \xi \leq x \leq \frac{\xi}{x_B}. \qquad (11)$$

This prescription has the nice feature that it naturally has an upper limit of integration that respects the physical limits. It has, however, a problem at the tree level, where they have a nonzero value for the structure functions at $x_B = 1$. In [5], they invoke jet mass corrections to solve this problem but it is, nevertheless, a phenomenological fix.

In addition to the above mentioned problem, there are two extra limitations to this prescription: (a) why should the baryon number flow entirely through the lower part of the graph, and (b) the first moment of the structure functions cannot be separated into a soft and a hard part [7].

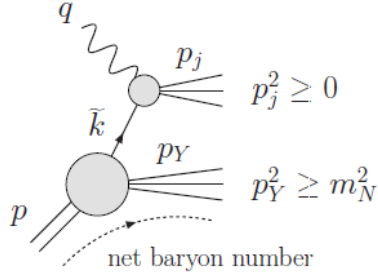

**FIGURE 1.** Graph for the interaction of the target with the incoming photon. Extracted from [5].

## CONCLUSIONS

The OPE approach to the target mass corrections has the side effect that the parton distributions are defined in an unphysical region. It is expected [2,6] that higher twists will cancel such unphysical contributions, but so far it is only an expectation. The partonic approach [3,6] in the transverse basis reproduces OPE, in its glory and in its failure. If one includes the correct physical contribution to the OPE, one then is led to the breakdown of the concept of universal parton distributions [4]. Even if one switches to a scheme involving collinear factorization [5], none of these problems are solved. In the end, it seems that if TMC are included, one loses the partonic interpretation: no parton distribution with TMC can be really defined. Given the importance of the potential consequences, it is urged that more investigation on this fascinating subject be made.

## ACKNOWLEDGMENTS

I would like to thank Wally Melnitchouk and Alberto Accardi for fruitful discussions, and the conference committee for partially supporting my visit. This work was supported by CNPq.